\documentclass[runningheads]{llncs}
\usepackage{graphicx}
\usepackage{cite}
\usepackage{amsmath,amssymb,amsfonts}
\usepackage{algorithmic}
\usepackage{graphicx}
\usepackage{textcomp}
\usepackage{xcolor}
\usepackage{multirow}
\usepackage{subcaption}
\usepackage{listings}
\usepackage{hyperref}
\usepackage{dirtree}

\definecolor{codegreen}{rgb}{0,0.6,0}
\definecolor{codegray}{rgb}{0.5,0.5,0.5}
\definecolor{codepurple}{rgb}{0.58,0,0.82}
\definecolor{backcolour}{rgb}{0.95,0.95,0.92}

\lstdefinestyle{mystyle}{
    backgroundcolor=\color{backcolour},   
    commentstyle=\color{codegreen},
    keywordstyle=\color{magenta},
    numberstyle=\tiny\color{codegray},
    stringstyle=\color{codepurple},
    basicstyle=\ttfamily\footnotesize,
    breakatwhitespace=false,         
    breaklines=true,                 
    captionpos=b,                    
    keepspaces=true,                 
    numbers=left,                    
    numbersep=5pt,                  
    showspaces=false,                
    showstringspaces=false,
    showtabs=false,                  
    tabsize=2
}

\def\BibTeX{{\rm B\kern-.05em{\sc i\kern-.025em b}\kern-.08em
    T\kern-.1667em\lower.7ex\hbox{E}\kern-.125emX}}

\newcommand{\ttilde}{{\raise.17ex\hbox{$\scriptstyle\sim$}} }

\begin{document}

\title{MIST: A Simple and Scalable End-To-End 3D Medical Imaging Segmentation Framework}
\titlerunning{MIST}
\author{
Adrian Celaya\inst{1,2} \and
Evan Lim\inst{1} \and
Rachel Glenn\inst{1} \and
Brayden Mi\inst{1} \and
Alex Balsells\inst{1,2} \and
Dawid Schellingerhout\inst{1} \and
Tucker Netherton\inst{1} \and
Caroline Chung\inst{1} \and
Beatrice Riviere\inst{2} \and
David Fuentes\inst{1}
}
\authorrunning{A. Celaya et al.}

\institute{
The University of Texas MD Anderson Cancer Center, Houston TX 77030 \and
Rice University, Houston TX 77005
}

\maketitle 
\begin{abstract}
Medical imaging segmentation is a highly active area of research, with deep learning-based methods achieving state-of-the-art results in several benchmarks. However, the lack of standardized tools for training, testing, and evaluating new methods makes the comparison of methods difficult. To address this, we introduce the Medical Imaging Segmentation Toolkit (MIST), a simple, modular, and end-to-end medical imaging segmentation framework designed to facilitate consistent training, testing, and evaluation of deep learning-based medical imaging segmentation methods. MIST standardizes data analysis, preprocessing, and evaluation pipelines, accommodating multiple architectures and loss functions. This standardization ensures reproducible and fair comparisons across different methods. We detail MIST's data format requirements, pipelines, and auxiliary features and demonstrate its efficacy using the BraTS Adult Glioma Post-Treatment Challenge dataset. Our results highlight MIST's ability to produce accurate segmentation masks and its scalability across multiple GPUs, showcasing its potential as a powerful tool for future medical imaging research and development.
\end{abstract}

\section{Introduction}
Medical imaging segmentation is a highly active area of research. Since its introduction in 2015, the U-Net architecture has received nearly 90,000 citations on Google Scholar \cite{unet, unet-3d}. Other architectures and frameworks like nnUNet have achieved state-of-the-art accuracy in several medical imaging benchmarks like the Brain and Tumor Segmentation (BraTS) and Medical Segmentation Decathlon (MSD) \cite{isensee2021nnu, isensee2021nnubrats, antonelli2022medical}. Since nnUNet's introduction in 2018, several innovative approaches for medical imaging segmentation have emerged. These innovations include new architectures like vision transformers \cite{vaswani2017attention} and loss functions like the boundary and generalized surface losses \cite{bl, celaya2023generalized}. Despite these recent advances, there remains a lack of standardized tools for testing and evaluating different approaches for medical imaging segmentation. For example, several works like \cite{tang2022self, chen2021transunet, hatamizadeh2021swin, zhou2021nnformer, cao2022swin} report superior performance to nnUNet while \cite{isensee2024nnurevisit} refutes these results. This inconsistency makes it difficult to evaluate and assess claims of state-of-the-art performance for new research in deep learning-based medical imaging segmentation.

To address this inconsistency, we propose the Medical Imaging Segmentation Toolkit or MIST, a simple, modular, and end-to-end deep learning-based medical imaging segmentation framework that allows researchers to seamlessly train, test, and evaluate existing and new methods on their data in a consistent and reproducible way. MIST has a rule-based data analysis and preprocessing pipeline that it makes available to multiple architectures, loss functions, and training parameters. Additionally, MIST has a standardized and flexible evaluation pipeline that computes multiple metrics like the Dice, Hausdorff distance, and surface Dice for any user-defined segmentation class. This standardization of several pipelines allows for fair comparisons between methods and their effect on segmentation accuracy for any given dataset.

\section{Methods} \label{sec:methods}
In this section, we describe the required data format for MIST, the different pipelines in MIST, and the auxiliary features, such as evaluation, postprocessing, and test-time inference. We also provide the specific settings we use to train on the BraTS challenge data.

\subsection{Data Format} \label{sec:methods-data-format}
MIST uses the same file structure as the BraTS challenge datasets \cite{de20242024, karargyris2023federated}. That is, MIST assumes that the train and test (optional) data directories are formatted so that each sub-directory corresponds to one patient and that each imaging modality (i.e., T1, T2, or CT) is its own NIfTI file. The labels should all be in a single NIfTI file in the same sub-directory for each patient. Once the dataset is in the correct format, the next step is to prepare a small JSON file containing the details of the dataset. These include information like the imaging modality, naming conventions for the images and masks, labels, and the final classes that we use for evaluation (i.e., labels 1, 2, and 3 for whole tumor). An example of this JSON file is provided in the documentation page \href{https://mist-medical.readthedocs.io/en/latest/getting_started/}{here}. MIST also provides support for converting datasets that are in CSV or MSD format. 

\subsection{Pipelines} \label{sec:methods-pipelines}
MIST contains three main pipelines: Data Analysis, Preprocessing, and Training. These pipelines can all run at once with a single command or individually. We also provide several auxiliary pipelines for evaluation, postprocessing, and test-time inference.

\subsubsection{Data Analysis} 
The data analysis pipeline gathers parameters about the dataset including cropping to foreground, target spacing, patch size selection, normalization parameters, and is described further below. The results of this analysis are saved in a \verb|config.json| file.

\paragraph{Cropping to Foreground} Often, zeros or useless background information surround the region of interest. For example, zeros surround the region outside of the brain in the BraTS datasets, and the table and air surrounding the body in CT imaging do not provide relevant information. From a computational perspective, it is desirable to crop the images to the foreground to make them smaller. Additionally, cropping to the foreground also reduces the sparsity of the dataset. MIST crops each image by first windowing it with its 33rd and 99.5 percentile values, then using an Otsu filter to produce a foreground threshold. Note that the windowing percentiles were determined experimentally with several datasets like LiTS, MSD, NFBS, BraTS, and AMOS \cite{lits, nfbs, ji2022amos}. We then crop the image according to the binary mask resulting from this threshold (i.e., one if the voxel intensity is greater or equal to the threshold). To determine if we need to use cropping, we check to see if, on average, the resulting volume reduction is at least 20\%.

\paragraph{Target Spacing} To determine the target spacing for a dataset, we first look at the distribution of the voxel spacings along each axis. Our initial target spacing is the median spacing in each direction. However, suppose the resulting target spacing is anisotropic (i.e., the maximum and minimum spacing ratio is greater than three), then we take the 10th percentile spacing along the lowest resolution axis in that case. However, if resampling with this selected target spacing results in an example taking up more than 2 GB of memory, then MIST will print a warning to users asking them to consider coarsening the image. We set this 2 GB threshold to conserve computational resources. Larger images take longer to load from disk and require more resources to manipulate (i.e., patch extraction and augmentation).

\paragraph{Patch Size Selection} We set the patch size for training from the median resampled image size by taking the nearest power of two less than or equal to each dimension in the median image size up to a maximum patch size. The default value for this maximum patch size is 256$\times$256$\times$256. However, users can adjust this maximum patch size based on their knowledge of their computing resources. 

\paragraph{Normalization Parameters} If the dataset consists of MR or other non-CT images, then our windowing parameters are the 0.5 and 99.5 percentile values of either the entire image or the non-zero values of the image if, on average, the ratio of the number of non-zero to zero-valued voxels is less than 0.2. We use the mean and standard deviation with the same rules regarding the non-zero values for normalization. On the other hand, if we are dealing with CT images, then we compute the 0.5 and 99.5 percentile values, mean, and standard deviation of all of the voxels over the entire dataset corresponding to regions labeled as non-zero in the ground truth segmentation masks.

\paragraph{Other Checks} In addition to the previous analysis steps, MIST checks to see that the header information in the images and masks agree and, if there are multiple modalities, that the headers in each image agree. If there is a disagreement between the headers, MIST prints a warning with the patient ID to the console and excludes the example from training.

\subsubsection{Preprocessing} 
\begin{figure}[ht!]
    \centering
    \includegraphics[width=\linewidth]{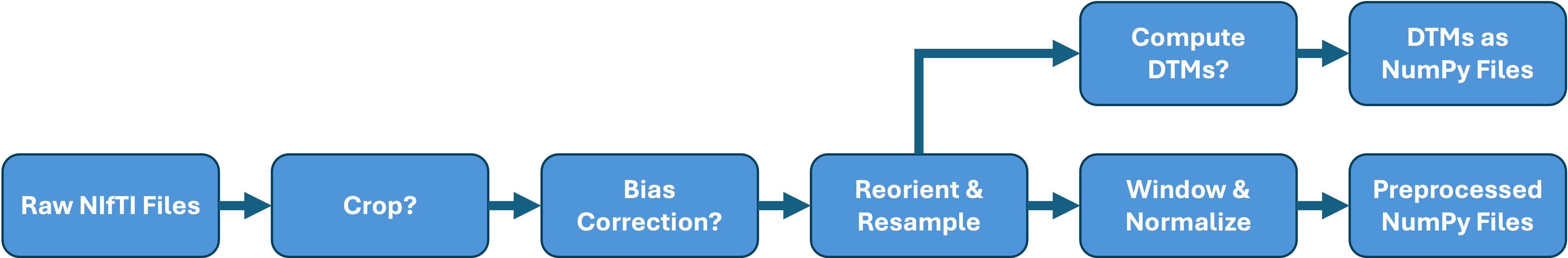}
    \caption{Workflow for the MIST preprocessing pipeline. This preprocessing pipeline takes raw NIfTI files and outputs cropped (optional), bias corrected (optional and MR only), reoriented, resampled, windowed, and normalized NumPy files. This pipeline can also compute the DTMs for each label in the ground truth mask and output them as NumPy files. \label{fig:mist-preprocessing}}
\end{figure}

The preprocessing pipeline takes the information from the analysis pipeline to crop the images based on their foreground mask (if called for), reorient to right-anterior-inferior (RAI), resample to the target spacing, and window and normalize the intensity values. This pipeline can also compute the distance transform maps (DTMs) for each class in the ground truth masks or apply bias correction for MR images. Users also have the option to skip this pipeline if, for example, their NIfTI files contain preprocessed images already. Figure \ref{fig:mist-preprocessing} illustrates the workflow for this pipeline.

\paragraph{Reorient \& Resample} After optionally cropping or bias correcting each image, the preprocessing pipeline will reorient each image and mask to the RAI orientation. For resampling images to the target spacing, we use third-order spline interpolation. We apply one-hot encoding and use linear interpolation to each channel for masks. If the current image spacing is anisotropic, we first resample along the low-resolution axis using nearest-neighbor interpolation and then apply the appropriate interpolation along the other two axes \cite{isensee2021nnu}. We implement these operations using SimpleITK to optimize the speed and computational performance of these operations \cite{lowekamp2013design}. 

\paragraph{Window \& Normalize} Before normalizing the voxel intensities in each image, the preprocessing pipeline windows each image by either the image's 0.5 and 99.5 percentile or, if we are using CT images, by the precomputed windowing parameters. After windowing, MIST applies z-score normalization with either the image's mean and standard deviation or the precomputed normalization parameters for CT images. Again, in the case that the ratio of non-zero to zero-valued voxels is less than 0.2, then the mean and standard deviation are computed from the non-zero voxels only, and we multiply the resulting images by the non-zero mask (different from foreground mask) to preserve the zero-valued voxels.

\paragraph{Distance Transform Maps} A DTM represents the distance between each voxel and an object's closest boundary or edge. The values in a DTM are positive on the exterior, zero on the boundary, or negative in the object's interior. As a result, they can provide helpful information during training. Several loss functions and shape-aware networks use DTMs to improve the accuracy of medical imaging segmentation methods \cite{bl, celaya2023generalized, karimi2019reducing, li2020shape}. MIST allows users to pre-compute the DTMs for each label in the ground truth segmentation masks. Each channel in the resulting DTM represents the DTM for that label (i.e., channel one is the DTM for label one). If a label does not exist in the image, the DTM will be a 3D array with all values being the diagonal distance of the image. Additionally, MIST allows users to output normalized DTMs whose values range from -1 to 1. 

\subsubsection{Training}
Once preprocessing is complete, MIST uses a five-fold cross-validation to train and evaluate a model on the dataset. In other words, for each fold MIST sets aside 20\% of the data as a validation dataset and uses the remaining 80\% of the data for training. Users can adjust these percentages, specify a different number of folds, or even specify custom folds if, for example, they are running leave-one-institution-out cross-validation.

MIST's default behavior is to use all available GPUs. When MIST runs multi-GPU training, it uses data parallelism via PyTorch's DistributedDataParallel package \cite{li2020pytorch}. Additionally, MIST uses the Nvidia Data Loading Library (DALI) to handle data loading, patch extraction, and random augmentation. This feature further optimizes its computational performance by running most of the operations on the GPUs. 

\paragraph{Network Architectures} MIST supports six different network architectures. These architectures include nnUNet (default) \cite{isensee2021nnu}, U-Net \cite{unet, unet-3d}, Swin UNETR \cite{cao2022swin}, PocketNet \cite{celaya2022pocketnet}, and MedNeXt \cite{roy2023mednext}. MIST also provides users with options like adding deep supervision \cite{wang2015training} or variational autoencoder regularization \cite{tang2021variational} to most of these architectures. Other regularization features like L2 and L1 regularization and gradient clipping are available for all architectures. For more details about the supported architectures (and their differences), please refer to our \href{https://mist-medical.readthedocs.io/en/latest/advanced_topics/#network-architectures}{documentation page}. Finally, MIST's modular design allows users to implement their own architectures and integrate them into MIST with minimal effort. 

\paragraph{Loss Functions} Like with the choice of network architecture, MIST allows users to choose from a variety of loss functions like Dice with Cross Entropy (default) \cite{milletari2016v}, the clDice loss \cite{shit2021cldice}, and boundary-based loss functions like the Boundary, Hausdorff, and Generalized Surface losses \cite{bl, karimi2019reducing, celaya2023generalized}. Again, MIST's modular design allows researchers to implement custom loss functions and seamlessly use them within the MIST framework. 

\paragraph{Other Training Options} Finally, MIST gives users the flexibility to choose different optimizers like Adam (default) \cite{adam}, stochastic gradient descent, and AdamW \cite{loshchilov2017decoupled}. MIST also supports several learning rate schedulers, with a constant learning rate of 0.0003 being the default. Additional features including transfer learning are included in our documentation page \href{https://mist-medical.readthedocs.io/en/latest/}{here}.

\subsubsection{Auxiliary Pipelines}

\paragraph{Evaluation} Once training is complete, the evaluation pipeline launches to assess the accuracy of the predictions with respect to the final classes defined by the user in their dataset JSON file. By default, these metrics are the Dice similarity coefficient and the 95th percentile Hausdorff distance (HD95). Other metrics like the surface Dice and average surface distance are also available. The output of the evaluation pipeline is a CSV file called \verb|results.csv| where each row shows the accuracy for each class and metric for one patient. The last five rows give the mean, standard deviation, median, and 25th and 75th percentiles for all of the metrics and classes. The evaluation pipeline is also available as a stand-alone pipeline that can run on any set of predictions. With the exception of the surface Dice, all of our metrics are in line with \href{https://metrics-reloaded.dkfz.de/}{Metrics Reloaded} \cite{maier2024metrics}.

\paragraph{Postprocessing} Postprocessing is available as an optional stand-alone pipeline. This pipeline allows users to specify which labels to apply a user-specified set of operations. These operations include the removal of small objects, only keeping the top-$k$ components (adjustable value of $k$), morphological cleaning, and filling holes with a specified value. Once the postprocessing pipeline finishes, the evaluation pipeline launches and compares the change in accuracy to the baseline results. MIST computes this change in accuracy as a weighted average of the specified metrics. Note that MIST applies postprocessing to the discrete predictions, not the softmax probabilities.

\paragraph{Test-Time Inference} Finally, MIST provides a stand-alone inference pipeline that takes as input one or more MIST models and a CSV or JSON file with the paths to test-time data. This pipeline writes test-time predictions to a user-specified output directory. 

\subsection{BraTS 2024 Training Protocols} \label{sec:methods-brats}
Our choice of architecture is the Pocket nnUNet with deep supervision (two supervision heads) and with residual convolution blocks \cite{celaya2022pocketnet, isensee2021nnu}. MIST automatically selects a patch size of 128$\times$128$\times$128. We use eight NVIDIA H100 GPUs with a batch size of 32 uniformly distributed across the GPUs. Additionally, we use L2 regularization with a penalty parameter equal to 1e-5. Our choice of loss function is the Dice with Cross Entropy loss. While the default number of epochs per fold for MIST is 1,000, we train for 15,000 epochs per fold. We use a cosine learning rate schedule with an initial learning rate of 0.001 \cite{loshchilov2016sgdr}. Within each fold, we set aside 2.5\% of the training data (27 patients) as a validation set to select the best model. Once training is complete, MIST runs inference on the challenge validation data using test time augmentation (flipping along each axis and averaging each prediction) for each model. Inference uses sliding windows with an overlap of 0.5 with Gaussian blending ($\sigma=0.125$). The predictions from all five models are averaged to produce a final prediction. All other options and hyperparameters are left at their default values. Please refer to MIST's \href{https://mist-medical.readthedocs.io/en/latest/}{documentation} for these values.

\section{Results} \label{sec:results}
\subsection{Accuracy}
We use MIST and the parameters described in Section \ref{sec:methods-brats} to perform a five-fold cross-validation (CV) with the BraTS Adult Glioma Post-Treatment Challenge data \cite{de20242024}. Table \ref{tab:results-train} gives the results of this five-fold CV in terms of the Dice score and HD95. Tables \ref{tab:results-validation} and \ref{tab:results-validation-lesion-wise} give the accuracy of the predictions in the validation set in terms of the same metrics and their lesion-wise versions.

\begin{table}[ht!]
\centering
\setlength{\tabcolsep}{12pt}
\resizebox{\linewidth}{!}{%
\begin{tabular}{lccccc}
\hline
\multicolumn{1}{c}{\multirow{2}{*}{Class}} & \multicolumn{1}{l}{\multirow{2}{*}{Labels}} & \multicolumn{2}{c}{Dice} & \multicolumn{2}{c}{HD95 (mm)} \\
\multicolumn{1}{c}{} & \multicolumn{1}{l}{} & Mean (Std.) & Median & Mean (Std.) & Median \\ \hline
NETC & 1 & 0.8068 (0.3173) & 1.0000 & 27.492 (87.446) & 0.0000 \\
SNFH & 2 & 0.8978 (0.1027) & 0.9262 & 5.0248 (15.888) & 1.8660 \\
ET & 3 & 0.8030 (0.2823) & 0.9193 & 26.638 (84.388) & 1.0000 \\
RC & 4 & 0.7754 (0.2994) & 0.9066 & 30.955 (88.628) & 2.2361 \\
TC & 1, 3 & 0.7962 (0.2870) & 0.9183 & 27.983 (85.971) & 1.7321 \\
WT & 1, 2, 3 & 0.9063 (0.0100) & 0.9328 & 5.0210 (15.846) & 2.0000 \\ \hline
\end{tabular}%
}
\vspace{2mm}
\caption{Dice and 95th percentile Hausdorff distances for each segmentation class after a five-fold cross-validation on the BraTS Adult Glioma Post-Treatment Challenge training data.}
\label{tab:results-train}
\end{table}

\begin{table}[ht!]
\centering
\setlength{\tabcolsep}{27pt}
\resizebox{\linewidth}{!}{%
\begin{tabular}{lccc}
\hline
\multicolumn{1}{c}{Class} & \multicolumn{1}{l}{Labels} & Dice & HD95 (mm) \\ \hline
NETC & 1 & 0.7010 & 58.165 \\
SNFH & 2 & 0.9208 & 6.6536 \\
ET & 3 & 0.7427 & 27.950 \\
RC & 4 & 0.6841 & 49.321 \\
TC & 1, 3 & 0.7379 & 28.514 \\
WT & 1, 2, 3 & 0.9257 & 6.6367 \\ \hline
\end{tabular}%
}
\vspace{2mm}
\caption{Validation accuracy in terms of the Dice and 95th percentile Hausdorff distances for each segmentation class in the BraTS Adult Glioma Post-Treatment Challenge validation data. Note that we only report the mean for each metric here since the competition evaluation pipeline does not provide the other statistics in this case.}
\label{tab:results-validation}
\end{table}

\begin{table}[ht!]
\centering
\setlength{\tabcolsep}{12pt}
\resizebox{\linewidth}{!}{%
\begin{tabular}{lccccc}
\hline
\multicolumn{1}{c}{\multirow{2}{*}{Class}} & \multicolumn{1}{l}{\multirow{2}{*}{Labels}} & \multicolumn{2}{c}{Lesion-Wise Dice} & \multicolumn{2}{c}{Lesion-Wise HD95 (mm)} \\
\multicolumn{1}{c}{} & \multicolumn{1}{l}{} & Mean (Std.) & Median & Mean (Std.) & Median \\ \hline
NETC & 1 & 0.7561 (0.3711) & 1.0000 & 53.029 (115.98) & 0.0000 \\
SNFH & 2 & 0.8504 (0.2063) & 0.9449 & 30.391 (67.397) & 1.4142 \\
ET & 3 & 0.7253 (0.3195) & 0.8624 & 46.885 (100.89) & 2.0000 \\
RC & 4 & 0.7024 (0.3595) & 0.8834 & 54.809 (116.18) & 3.5821 \\
TC & 1, 3 & 0.7066 (0.3147) & 0.8361 & 49.134 (98.166) & 3.5056 \\
WT & 1, 2, 3 & 0.8602 (0.1982) & 0.9451 & 29.309 (66.096) & 1.9126 \\ \hline
\end{tabular}%
}
\vspace{2mm}
\caption{Lesion-wise validation accuracy in terms of the Dice and 95th percentile Hausdorff distances for each segmentation class in the BraTS Adult Glioma Post-Treatment Challenge validation data.}
\label{tab:results-validation-lesion-wise}
\end{table}

\subsection{Computational Performance}
\begin{figure}[ht!]
    \centering
    \includegraphics[width=0.9\linewidth]{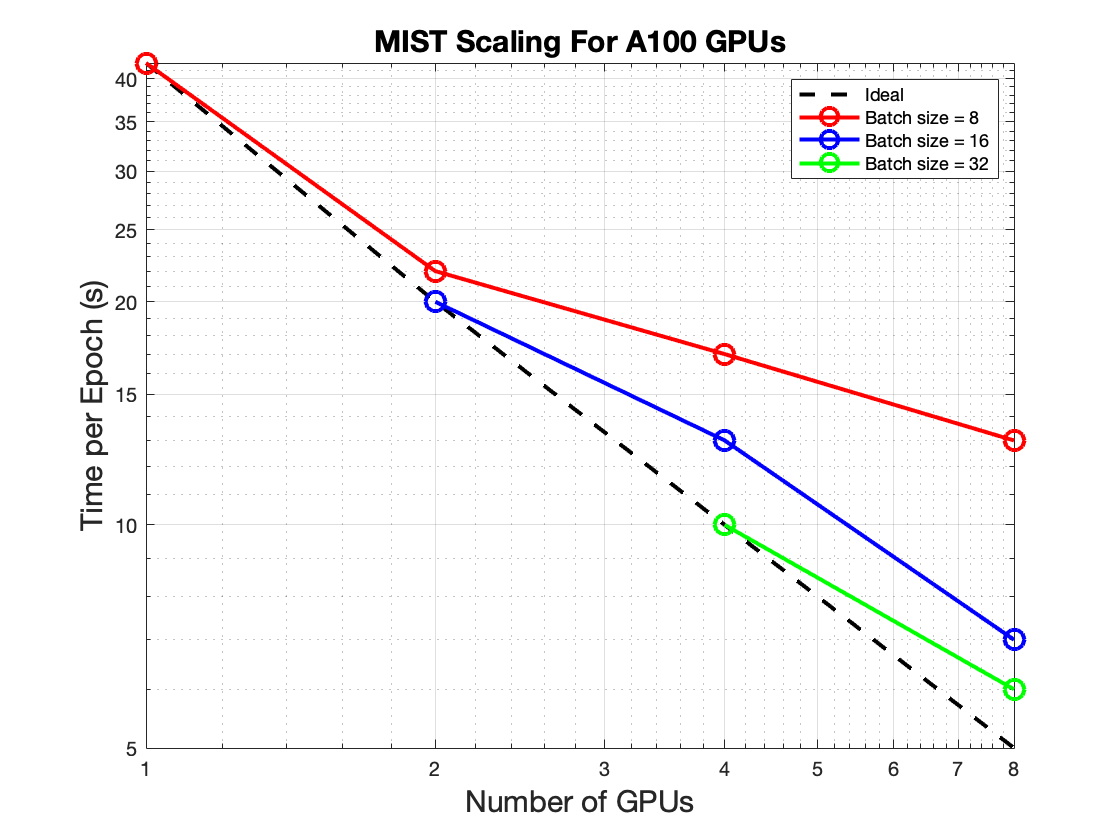}
    \caption{Time per epoch for different batch sizes and numbers of A100 GPUs. Here, we see that MIST achieves better scaling with larger batch sizes for A100 GPUs.\label{fig:a100-scaling}}
\end{figure}

\begin{figure}[ht!]
    \centering
    \includegraphics[width=0.9\linewidth]{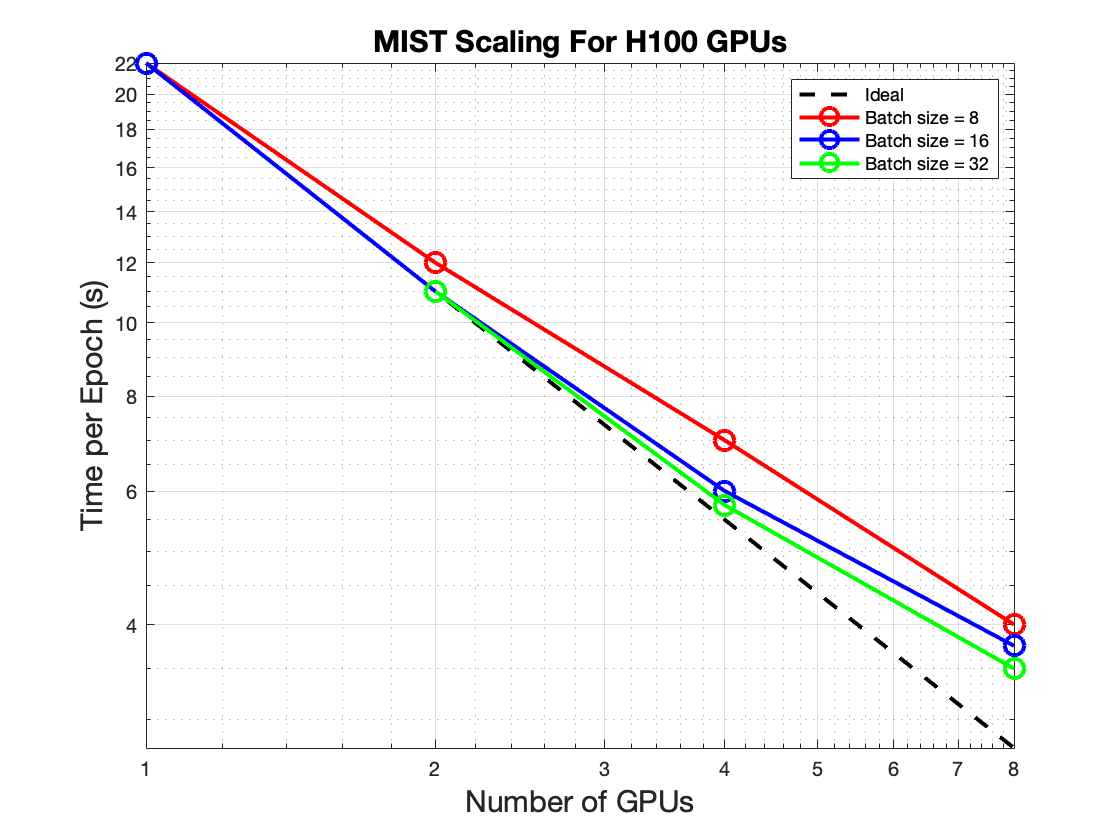}
    \caption{Time per epoch for different batch sizes and numbers of H100 GPUs. Here, we see that MIST achieves near optimal scaling for all batch sizes.\label{fig:h100-scaling}}
\end{figure}

In addition to the accuracy of our predictions, we also report the scalability of MIST for different numbers of A100 and H100 GPUs for multiple batch sizes. Figures \ref{fig:a100-scaling} and \ref{fig:h100-scaling} show these results for A100 and H100 GPUs, respectively. In these figures, we see that MIST effectively scales for multiple GPUs. This scaling is especially evident for the H100 GPUs. It is also important to note that we observe a roughly two times speed up with the H100 GPUs, which is in line with what Nvidia reports in their comparisons between the two types of graphics cards. 

\section{Discussion}
Our results indicate that models trained using MIST can produce accurate adult post-operative gliomas, achieving median Dice scores of at least 0.9 for all of the segmentation classes in our five-fold CV. We choose to submit the raw output of our MIST pipeline to the BraTS 2024 competition, omitting the use of post-processing. However, future submissions to the continuous evaluation and competitions will explore different post-processing strategies with MIST by using its built-in post-processing pipeline. Indeed, previous BraTS competitions cite replacing small objects belonging to the ET class with the SNFH label as beneficial \cite{isensee2021nnubrats}. Additionally, we may see improvements in accuracy by using information from DTMs using boundary-based losses like the Generalized Surface Loss. Finally, different architectures like the multi-grid inspired FMG-Net or vision transformer-based architectures like Swin UNETR are also available on MIST \cite{celaya2023fmg, cao2022swin}. These architectures may also improve our results.

Future work will include an ablation study of the parameters that we selected for the BraTS dataset, like the use of deep supervision, residual convolution blocks, L2 regularization, and pocket architectures. Additionally, an in-depth comparison between other frameworks like nnUNet, MedNeXt, and Swin UNETR for this dataset will be an objective of future work. For an in-depth comparison of these frameworks for a liver segmentation dataset, please refer to \cite{patel2024training}.

MIST is able to utilize multiple GPUs to speed up training efficiently. This efficient use of multiple GPUs is especially true for H100s, where we see a nearly six times speed up when going from one to eight GPUs. The ability to take advantage of multiple GPUs and compute nodes effectively will be essential to keep training times manageable as larger datasets like TotalSegmentor become available \cite{wasserthal2023totalsegmentator, d2024totalsegmentator}. As the size and quality of publicly available 3D medical imaging datasets increase, the creation of foundation models for medical imaging segmentation will be beneficial to speed up training through transfer learning \cite{lee2024foundation, wang2023mis}. MIST already supports transfer learning with other MIST models. Given this capability and its scalability, MIST can be instrumental in the creation of future foundation models.

In conclusion, our results show that MIST is capable of achieving accurate segmentation results while also scaling well to multiple GPUs by effectively taking advantage of technologies like PyTorch's DDP and Nvidia's DALI loader. MIST's simplicity, accuracy, and scalability have the potential to make MIST a valuable tool in the development of foundation models for medical imaging tasks like segmentation and radiation treatment planning. MIST's modular design makes it easy for researchers to add new loss functions or architectures to its pipelines. This modularity and easy integration into MIST's existing pipelines allow for new research to be evaluated and compared to existing methods fairly and consistently. MIST is still under active development. We encourage researchers and the broader medical imaging community to contribute to its development by using it, providing feedback, or contributing new features. 

\section{Availability \& Attribution}
MIST is an open-source package (Apache 2.0 license) and is available on \href{https://github.com/mist-medical/MIST}{GitHub} or \href{https://pypi.org/project/mist-medical/}{PyPI}. Please cite this manuscript, \cite{celaya2022pocketnet}, and \cite{celaya2023fmg} if you use MIST for your own publications.

\section{Acknowledgments}
The Department of Defense supports Adrian Celaya through the National Defense Science \& Engineering Graduate Fellowship Program. This research was partially supported by the Tumor Measurement Initiative through the MD Anderson Strategic Research Initiative Development (STRIDE), NSF-2111147, NSF-2111459, and NIH R01CA195524

\bibliographystyle{splncs04}
\bibliography{sources}

\end{document}